\pdfoutput=1

\documentclass[twocolumn,preprintnumbers,amsmath,amssymb,aps]{revtex4}

\usepackage{graphicx}
\usepackage{bm}
\usepackage[bookmarks]{hyperref}

\newcommand{\ket} [1] {\vert #1 \rangle}

\begin{document}

\title{Collinear source of polarization-entangled photon pairs\\at non-degenerate wavelengths}

\author{P. Trojek}
\author{H. Weinfurter}
 \affiliation{%
Department f\"{u}r Physik, Ludwig-Maximilians Universit\"{a}t, D-80799 M\"{u}nchen, Germany \\
Max-Planck-Institut f\"{u}r Quantenoptik, D-85748 Garching, Germany
}%

\date{January 12, 2008}

\begin{abstract}
We report on a simple but highly efficient source of polarization-entangled
photon pairs at non-degenerate wavelengths. The fully collinear configuration
of the source enables very high coupling efficiency into a single optical
mode and allows the use of long nonlinear crystals. With optimized dispersion
compensation it is possible to use a free-running laser diode as pump source
and to reach an entanglement fidelity of 99.4 \% at rates as high as 27000
pairs/s per mW of pump power. This greatly enhances the practicality of the
source for applications in quantum communication and metrology.
\end{abstract}

\maketitle

Entangled photons, for long time considered only as a tool for testing fundamental
aspects of quantum theory \cite{Freedman72}, now become a basic building block
for novel quantum communication and computation protocols, such as quantum
cryptography \cite{Ekert91}, dense coding \cite{Bennett92} or teleportation
\cite{Bennett93}. For the practical implementations, polarization encoding of
photons is among the best choices due to the availability of reliable
polarization-control elements and analyzers enabling high-fidelity measurements.
To date, the most established way to generate entangled photon pairs is
spontaneous parametric down-conversion (SPDC). In this process photons of an
intense pump beam spontaneously convert in a second-order ($\chi^{(2)}$)
nonlinear crystal into two lower-frequency photons provided that energy and
momentum is conserved. Two basic methods how to obtain polarization-entanglement
from SPDC are widely applied: the first uses type-II phase-matching in a single
crystal \cite{Kwiat95}, and the second relies on the coherent spatial overlap
of the emissions from two adjacent type-I phase-matched crystals \cite{Kwiat99}.
Usually, due to the non-collinear geometry of the methods, the nonlinear crystal
has to be relatively short (typically in range from 0.5 mm to 3 mm) and only a
small fraction of the emitted SPDC flux can be collected, thereby limiting
strongly the potential output brightness. Recently, a two-way collinear emission
was employed inside a Sagnac interferometer to produce entangled photon pairs
with long crystals or highly non-linear glass fibers \cite{Wong06}. This increases
the output brightness significantly, but now requires interferometric alignment.

Here, we present a source of polarization-entangled photon pairs overcoming
many deficiencies of the prior art, including those outlined above. We utilize
the fact that photon pairs at non-degenerate wavelengths can be generated
collinearly with respect to the pump light by SPDC and can thus be collected
very efficiently into a single-mode fiber. The spectral information is then
exploited to split the photons into two distinct spatial modes using a wavelength
division multiplexer (WDM). This method produces polarization entanglement
directly, without a need for any post-selection\cite{Kiess93} or beam overlap
\cite{Wong06}, and can be used in all applications where the photons are observed
separately.
	
For the actual generation of the photons, SPDC in the simple two-crystal geometry
is applied \cite{Kwiat99}. Consider two adjacent nonlinear crystals, both
operated in type-I phase-matching configuration and pumped with linearly
polarized light. The orientation of otherwise identical crystals is adjusted
such that the optic axes of the first and second crystal lie in the vertical
and horizontal planes, respectively. If the polarization of the pump beam is
oriented under 45$^{\circ}$, SPDC occurs equally likely in either crystal,
producing pairs of horizontally- $(\ket{H}\ket{H})$ or vertically-polarized
$(\ket{V}\ket{V})$ photons. By angle tuning the crystals, down-conversion is
set to a collinear phase-matching configuration \cite{Kim01}, emitting pairs
of photons at non-degenerate wavelengths $\lambda_1$  and $\lambda_2$. Provided
that the two emission processes are coherent with one another, which is fulfilled
as long as there is no way of ascertaining whether a photon pair was produced in
the first or the second crystal, the entangled state 
$\ket{\phi(\varphi)}=1/\sqrt{2}(\ket{H}_{\lambda_1}\ket{H}_{\lambda_2} + 
e^{i \varphi} \ket{V}_{\lambda_1}\ket{V}_{\lambda_2})$  is produced.

\begin{figure}
\includegraphics{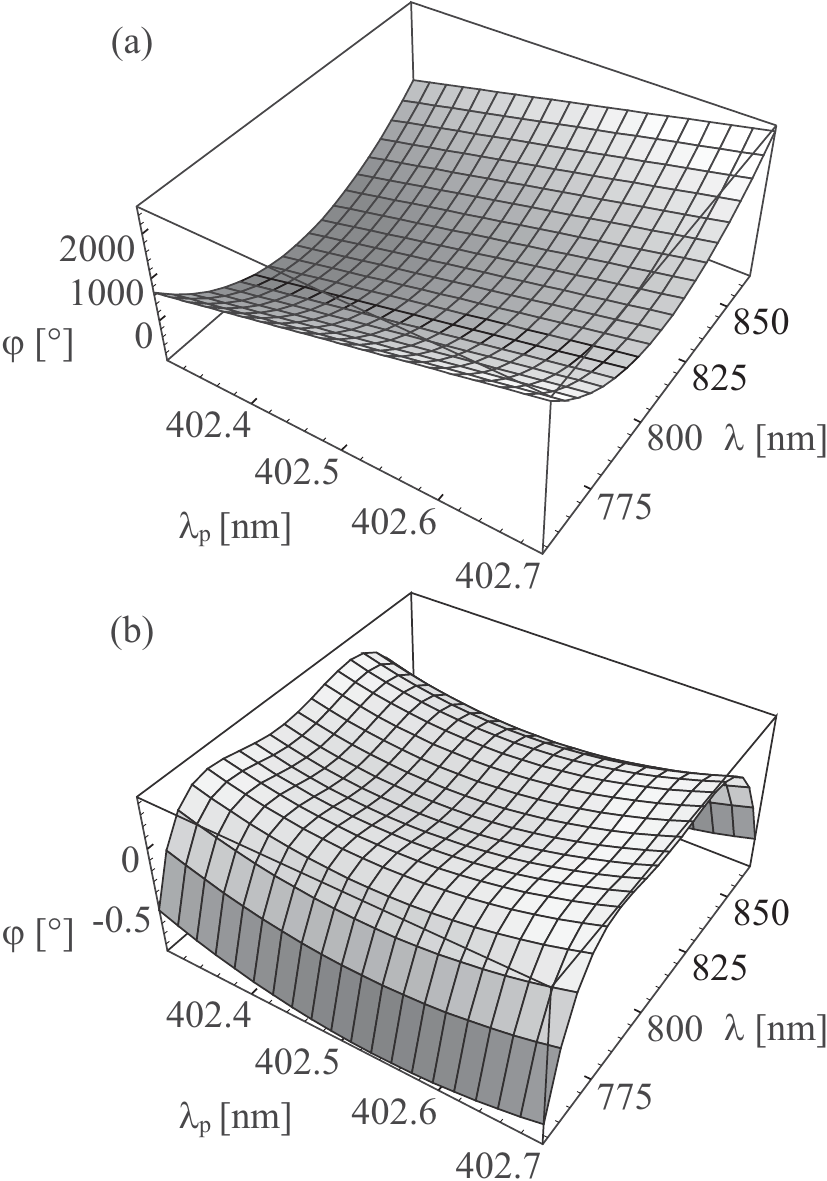}
\caption{\label{fig:phasemap} Calculated dependence of the relative phase $\varphi$ on
the wavelengths of pump $\lambda_p$  and one of the down-conversion photons
$\lambda$ for (a) uncompensated and (b) compensated configuration. The evaluation
assumes SPDC in the pair of BBO crystals, each 15.76 mm thick and cut at $\theta_p$
= 29.0$^{\circ}$; an overall phase offset is suppressed for clarity. After the
optimum compensation using a pair of tailored YVO$_4$ crystals with thicknesses of 8.20
and 9.03 mm, the phase surface is flat over the relevant spectral region (note the
change in the vertical scale), indicating a high purity of the entangled state.}
\end{figure}

In fact, dispersion and birefringence in the crystals lead to a partial loss of
coherence between the two emission processes. Because the times, when the photons
exit from the output face of the second crystal, depend on their wavelengths and
polarizations, they potentially reveal the actual position of the photon-pair
origin. This detrimental temporal effect is twofold, and can be better understood,
when inspecting the dependence of the relative phase $\varphi(\lambda_p,\lambda)$ on
the wavelengths of pump $(\lambda_p)$ and one of the down-conversion photons, see
Fig.~\ref{fig:phasemap}(a).
First, it is the group-velocity mismatch between the pump and the down-conversion
light, which causes that the photon pairs born in the first crystal are advanced
with regard to the pump photons and thus with regard to those originating in the
second crystal. This manifests itself as a finite slope of the phase map
$\varphi(\lambda_p,\lambda)$   in the  $\lambda_p$ direction and it is usually precluded
using narrowband pump \cite{Kwiat99}. To enable also the use of a broadband pump
source, e.g., a free-running laser diode, here we use a special birefringent
compensation crystal in the path of the pump, introducing a proper temporal
retardation between its horizontally- and vertically-polarized components
\cite{Kim01,Nambu02}, thus effectively pre-compensating the effect. Second, the
dispersive delay between the down-conversion photons at the non-degenerate
wavelengths is different for the two emission possibilities, because the photons
generated in the first crystal acquire an extra spread when propagating through
the second crystal. As a consequence, the relative phase $\varphi(\lambda_p,\lambda)$
varies strongly with the wavelength $\lambda$ as well. An additional birefringent
crystal with an effectively reverse phase characteristics has to be put behind the
SPDC crystals to counteract this second effect. In this way, the initially strongly
varying phase map $\varphi(\lambda_p,\lambda)$  becomes flat
[Fig.~\ref{fig:phasemap}(b)], indicating the complete temporal indistinguishability
of the emission processes. A more rigorous approach to the problem can be based on
the evaluation of photon time distributions \cite{Trojek07}.

\begin{figure}[b!]
\includegraphics{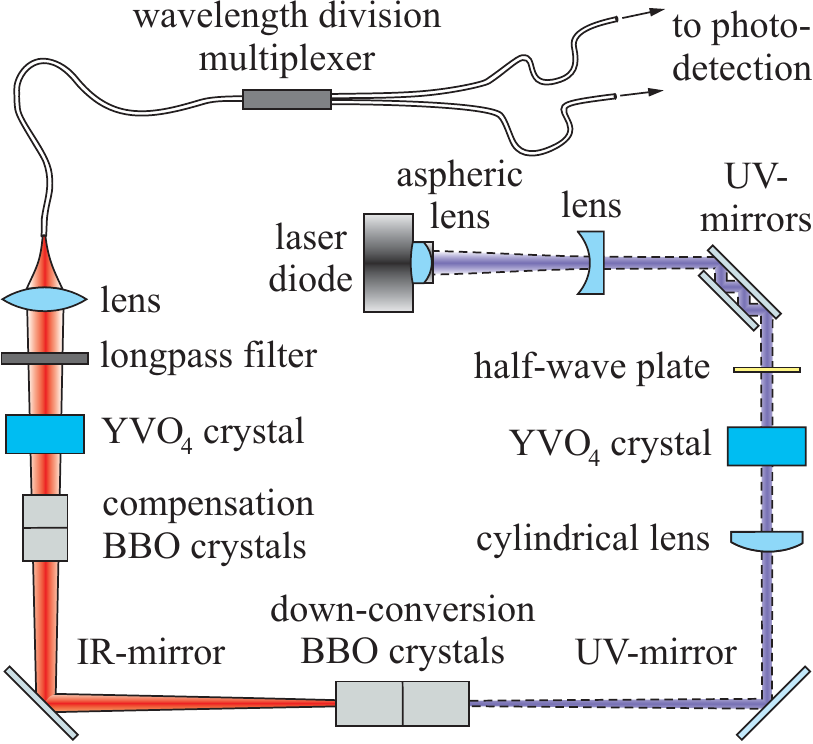}
\caption{\label{fig:setup} Scheme of the collinear source. A free-running violet
laser diode pumps a pair of BBO crystals and collinearly produces via SPDC pairs
of non-degenerate photons, which are collected into a single mode fiber. The
spectral information is exploited to separate the photons into two spatial modes
using a wavelength division multiplexer. Two compensation YVO$_4$ crystals and BBO
crystals are used to reverse the time delay effect and spatial lateral displacement
effect, respectively, introduced in the down-conversion BBO crystals.}
\end{figure}

The setup of the source is schematically shown in Fig.~\ref{fig:setup}.
The linearly polarized pump light is provided by a 60 mW free-running laser diode
operating at $\lambda_p = 403$ nm ($\Delta\lambda_p \approx 0.5$ nm). The light is
reflected several times at UV-mirrors to remove the broadband background laser
emission and passes through a half-wave plate rotating the polarization angle to
45$^{\circ}$. Three lenses are used to correct the astigmatism of the laser diode
emission and to focus the pump beam to a diameter of $\approx 220$ $\mu$m within
the two BBO ($\beta$-BaB$_2$O$_4$) crystals (each $6.0\times6.0\times15.76$ mm),
both cut for type-I phase matching at $\theta_p = 29.0^{\circ}$ and oriented for
a collinear emission of photons at the wavelengths of $\lambda_1 \approx 765$ nm
and $\lambda_2 \approx 850$ nm. The emitted photons are separated from the pump
light using an IR-mirror and a longpass filter, and are collected into a single-mode
fiber with an aspheric lens. To reach high collection efficiency, the lateral
displacement of the horizontally and the vertically polarized photons caused by
the birefringence of the BBO down-conversion crystals can be compensated best using
the same pair of BBO crystals, but only half as long. The photons are directly
guided to the custom made WDM, which splits the wavelengths $\lambda_1$ and
$\lambda_2$ with a probability of over 99\%. Labeling the spatial modes of the output
single-mode fibers by 1 and 2, the polarization state of photons becomes $\ket{\phi(\varphi)}=1/\sqrt{2}(\ket{H}_{1}\ket{H}_{2}+e^{i \varphi} \ket{V}_{1}\ket{V}_{2})$,
provided that the two emission alternatives are temporally indistinguishable.
This is achieved using two yttrium vanadate (YVO$_4$) crystals, with their thicknesses
$d$ optimized according to the criteria described above. The first one ($d = 8.20$ mm),
is put into the path of the pump, whereas the other ($d = 9.03$ mm) is included in the
path of the down-conversion photons. Both YVO$_4$ crystals are cut at 90$^{\circ}$ with
regard to pump- and fluorescence-light direction, in order to avoid spatial walk-off
effects. For the detection of the down-conversion photons two actively quenched silicon
avalanche photodiodes with separately measured efficiency of 50--51 \% at 800 nm were
used.

\begin{figure}
\includegraphics{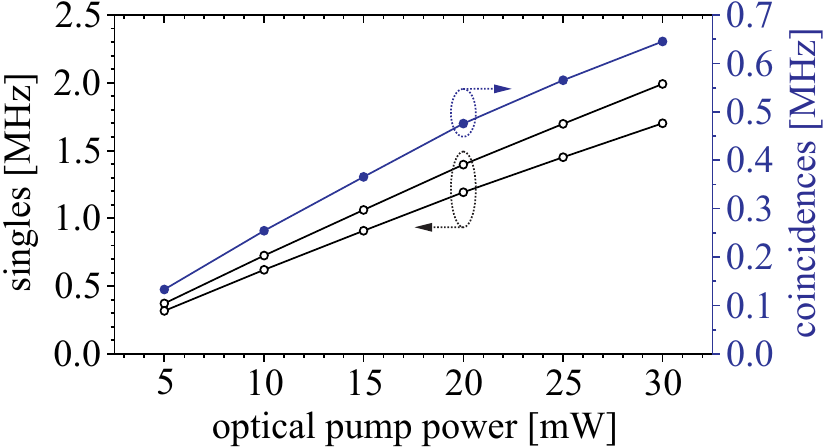}
\caption{\label{fig:rates} Detected single and coincidence count rates depending on
the pump power measured at the position of BBO crystals.}
\end{figure}

With this source, we detected 27000 pairs per second and milliwatt of pump power
(Fig.~\ref{fig:rates}).
Another important figure of merit, the coincidence-to-single ratio, reached values
between 0.36 and 0.39 depending on the pump power, thereby confirming the very high
coupling efficiency of photons into the single-mode fiber. Taking into account the
limited detection efficiency and other losses in the set-up, such as the reflection
at the fiber tips (all together $> 12$\%) and the optics in the path of down-conversion
photons ($> 3$\%) or the insertion loss of the WDM ($> 4$\%), we estimate the net
coupling efficiency to reach values as high as 90\%.

To verify the entanglement of photon pairs, the degree of polarization correlations
in two complementary bases was measured using a pair of polarizers. At low pump power
of $\approx 1$ mW we obtained a visibility of $V_{H/V} = 98.7 \pm 0.2$\% in the
horizontal/vertical basis and $V_{45} = 98.4 \pm 0.3$\% in the basis rotated by
45$^{\circ}$ or entanglement fidelity of $F = 99.4$\%, respectively. At higher pump
powers the increased single photon rates, together with a relatively long coincidence
gate time of $\tau = 5.8$ ns, make correction for accidental coincidences necessary.
The corrected visibilities are, within errors, consistent with those reported above.
The gap between the measured value and the maximum visibility of 1 is attributed to
depolarization inside the WDM.

\begin{figure}
\includegraphics{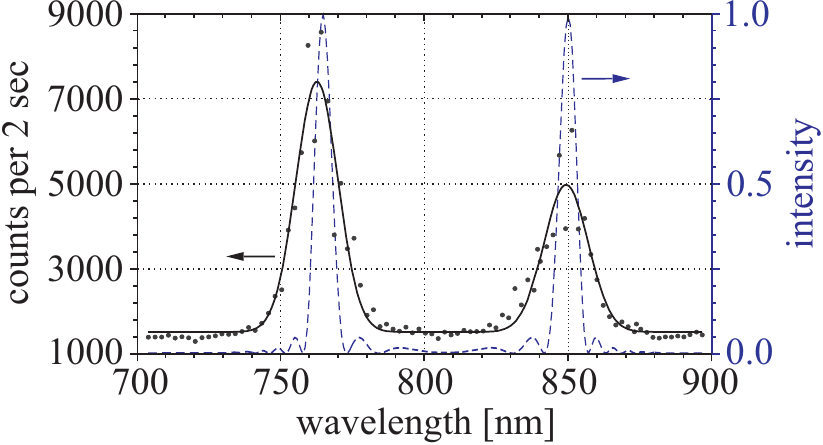}
\caption{\label{fig:spectrum} Spectral distribution of down-conversion light with the
central wavelengths of $\lambda_1 = 762.8 \pm 0.4$ nm and $\lambda_2 = 849.4 \pm 0.6$ nm,
determined from a Gaussian fit (solid line). The lower peak number of counts at $\lambda_2$ is
due to a reduced efficiency of the spectrometer (with resolution 1.2 nm) towards infrared
wavelengths. The dashed line shows the numerically simulated spectra obtained under the
assumption of narrowband pumping.}
\end{figure}

The spectral distribution of the collected down-conversion photons was measured to be
rather broad with the widths of  $\Delta\lambda_1 = 14.5 \pm 0.7$ nm and
$\Delta\lambda_2 = 15.4 \pm 1.2$ nm (Fig.~\ref{fig:spectrum}).
These values are in reasonable agreement with the theoretical values of
$\Delta\lambda_1 = 11.9$ nm and $\Delta\lambda_2 = 12.9$ nm obtained for the actual parameters
\cite{Trojek07}. Substantially reduced widths of $\Delta\lambda_1 = \Delta\lambda_2 = 6.4$ nm are
expected for narrowband pumping.

Finally, we note that in the current realization we could not take the expected advantage
of higher photon-pair fluxes converted in long down-conversion crystals. The additional
measurements with BBO crystals of lengths $L = 3.94$ mm and $L = 7.88$ mm reveal only a
negligible growth of the detected photon-pair rate $\propto L^{0.1}$. This dependence is
considerably slower than that inferred theoretically \cite{Ljunggren05}, $\propto \sqrt{L}$;
(for the scaling of the down-conversion bandwidth with $L$, there is a good agreement
between the experiment, $\Delta\lambda\propto L^{-0.68}$, and the theory,
$\Delta\lambda\propto L^{-0.73}$). We attribute the observed discrepancy to the
two-crystal geometry, which does not allow the optimum simultaneous coupling of photons
from both crystals, and to the spatial walk-off effect in the birefringent BBO crystals
resulting in a spatial asymmetry of the down-conversion emission.

In summary, the fully collinear configuration of the source utilizing only one spatial
mode for collecting down-conversion photons enables unprecedented coupling efficiencies
and brings many more advantages. First, it minimizes the complexity of the source and
thereby enhances its inherent robustness. Second, it precludes the occurrence of any
intrinsic spatial effect limiting the quantum-interference visibility, while at the same
time allows the use of long down-conversion crystals yielding higher photon-pair rates.
From a practical point of view, the technical requirements and the demand for alignment
are enormously reduced. These advantages make the source ideal for applications like
quantum cryptography \cite{Ekert91}, multiparty single qubit communication \cite{Schmid05}
and single photon detector calibration \cite{Klyshko80}. The already high brightness
reported here can be further improved with periodically poled crystals, allowing to
noncritically phase match any set of wavelengths and thus avoid the spatial walk-off
effects. In addition, it becomes possible to access the largest nonlinear coefficients,
which in turn suggests that an increase of photon-pair rates by 1--2 orders of magnitude
should be certainly feasible in the future.

The authors thank Y. Nazirizadeh for his assistance in the early stages of the project.
This work was supported by Deutsche Forschungsgemeinschaft through the DFG-Cluster of
Excellence MAP, by the DAAD-PPP Croatia program, and by the European Commission through
the EU projects QAP and SECOQC.

\end{document}